\documentclass{epl}

\title{Evidence for the coexistence of low-dimensional magnetism
and long-range order in \chem{Ca_3CoRhO_6}} \shorttitle{Evidence
for the coexistence of SRO and LRO in \chem{Ca_3CoRhO_6}}
\author{M. Loewenhaupt\inst{1} \and W. Sch\"{a}fer\inst{2} \and A.
Niazi\inst{3}  \and E.V. Sampathkumaran\inst{3}}
\institute{
  \inst{1} Institut f\"{u}r Festk\"{o}rperphysik, Technische Universit\"{a}t Dresden,
  01062 Dresden, Germany\\
  \inst{2} Mineralogisch-Petrologisches Institut, Universit\"{a}t Bonn, 53115 Bonn,
  Germany\\
  \inst{3} Tata Institute of Fundamental Research, Homi Bhabha Road,
  Mumbai-400 005, India}
\shortauthor{M. Loewenhaupt et al.}

\begin{document}

\maketitle

\begin{abstract}
We report the results of neutron powder diffraction studies on the
spin-chain compound \chem{Ca_3CoRhO_6} in the temperature range 3
to 293~K. Bragg peaks due to magnetic ordering start appearing
below about 100~K. The most interesting observation is that there
is a diffuse magnetic peak superimposed over the strongest
magnetic Bragg peak. The diffuse magnetic intensity is observed
below as well above 100~K. This finding provides a new insight
into the physics of this compound as though the low-dimensional
magnetic interaction coexists with long range magnetic order -- a
novel situation among quasi one-dimensional oxides.
\end{abstract}

\section{Introduction}
The pseudo low-dimensional insulators have been of constant
interest among physicists and chemists \cite{1} for the past few
decades as the interplay between intra-chain and inter-chain
interactions has been found to result in novel magnetic anomalies
\cite{2}. Till to-date, in all such compounds,  the presence of
weak inter-chain magnetic coupling results in eventual loss of the
identity of the chains, thereby paving the way for three
dimensional magnetic ordering at low temperatures. In this
article, we report the temperature dependent neutron diffraction
data for a quasi one-dimensional compound, \chem{Ca_3CoRhO_6},
which apparently has the signatures of the correlations within a
chain coexisting with long range magnetic order, thereby providing
a novel situation in this direction of research. The compound
under investigation crystallizes in the \chem{K_4CdCl_6}-type
rhombohedral structure (space group R\=3c). This class of
compounds, \chem{(Sr,Ca)_3MXO_6} ($M, X$ = a metallic ion,
magnetic or non-magnetic), have started attracting attention in
recent years due to the presence of spin-chains ($M-X$) separated
by $Sr/Ca$ ions (see refs. \cite{3} to \cite{16} and references
cited therein) and the magnetic chains form a triangular lattice
(with an inter-chain spacing of 5.313 \AA \ in the present
compound). In this structure, there is face-sharing of octahedra
of $X$ ions and trigonal prisms of $M$ ions. Of these, the title
compound is of special interest. On the basis of previous neutron
diffraction data \cite{16}, it was inferred that this compound can
be classified a "Partially Disordered Antiferromagnet (PDA)" -- a
novel magnetic structure very rarely encountered \cite{16} in
magnetism. In this magnetic structure, there is a temperature
range (below the paramagnetic state) in which the chains at the
apices of the hexagon are antiferromagnetically coupled to each
other, whereas the one at the center of the hexagon is left
incoherent; as the temperature is lowered further, the incoherent
chains can undergo spin-glass freezing (as proposed for the
present case \cite{16}) or couple ferrimagnetically with other
chains. It has been found that dc magnetic susceptibililty
($\chi$) exhibits a complex behavior at temperatures below 300~K
(refs. \cite{14} and \cite{15}), reflecting the existence of
different temperature regions: there is a broad maximum in the
plot of $\chi$ versus \textit{T} around 100 to 150~K, with an
isothermal magnetization behavior highly non-linear with the
magnetic field in the  intermediate \textit{T}-range (40 - 90~K),
but tending to a constant (zero-field-cooled) $\chi$ at still
lower temperatures. However, ac~susceptibililty  exhibits a broad
feature in the range 40 - 70~K with an unusually strong frequency
dependence, implying more exotic magnetic behavior of this
compound, thereby warranting further investigations for better
understanding \cite{15}. Motivated by this situation, we have
subjected this compound to more careful neutron diffraction
investigations as compared to those reported in ref. \cite{16},
the new results of which are reported in this article. It is
worthwhile stating that, in sharp contrast to the situation in
this material, the analogous \chem{Fe} compound exhibits a simple
magnetic behavior \cite{14} with a magnetic ordering temperature
around 15~K.

\section{Sample Preparation}
The polycrystalline specimen of \chem{Ca_3CoRhO_6} was synthesized
by a conventional solid state route. Stoichiometric amounts of
high purity ($>$99.9\%) \chem{CaCO_3}, \chem{CoO} and \chem{Rh}
powder were thoroughly mixed. Then the mixture was calcined at
900~C for one day. The prereacted powder was then finely ground,
pelletized and heated at 1200~C for about 10 days with few
intermediate grindings. The x-ray diffraction pattern confirmed
that the sample was single phase. Neutron powder diffraction
measurements were performed at the constant wavelength thermal
neutron instrument SV7 at the DIDO research reactor in the
Forschungszentrum J\"{u}lich \cite{17} using the diffractometers
SV7-a and SV7-b with neutron wavelengths of 1.095~\AA \ and
2.332~\AA, respectively. Both instruments are equipped with
JULIOS-type linear scintillation detectors. The sample was
contained in a vanadium cylinder of 8~mm diameter and 30~mm height
and inserted in a helium refrigerator cryostat equipped with
vanadium windows. Full long-term diffraction patterns were
collected at 293, 100, and 3~K in the short wavelength
configuration ($\lambda$ = 1.095~\AA) and at 4~K and 293~K using
the long wavelength of 2.332~\AA.  Short-term runs were performed
in the temperature regime between 3~K to 250~K for both
wavelengths.

\section{Results and discussion}

\begin{figure}
\onefigure{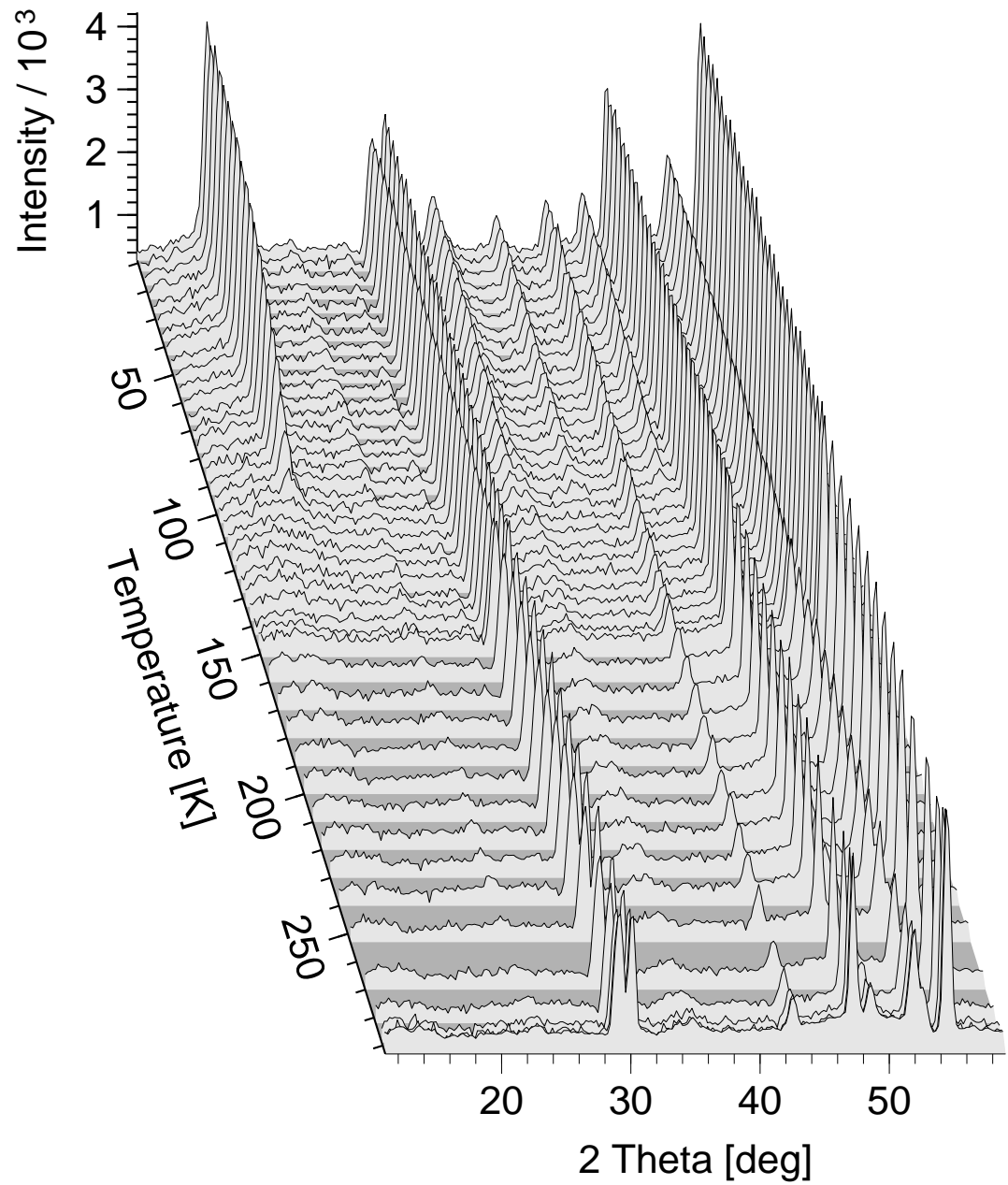} \caption{Temperature dependence of diffraction
patterns of \chem{Ca_3CoRhO_6} measured with a neutron wavelength
of 2.332~\AA.} \label{f.1}
\end{figure}

We show in fig.~\ref{f.1} the temperature dependence of the raw
diffraction patterns in the low-angle region with the long
wavelength. We have carried out crystal structure data analysis by
full-pattern Rietveld refinements using Fullprof ~\cite{18}.  A
two-step data handling was applied to the magnetic structure
analysis using (1) a modified version of the program Profan~
\cite{19} and (2) the full-matrix least squares program IC-POWLS~
\cite{20}. Profan fits pre-selectable profile functions,
\textit{e.g.} Gaussian or Lorentzian, into measured peaks or peak
clusters and individually refines sets of three profile parameters
each for peak position $2\Theta$, half width \textit{FWHM}, and
peak height~\textit{H}. This procedure permits the separation of
broad diffuse magnetic scattering contributions from the pure
Bragg peaks and thus the separation of disorder and long-range
order effects, respectively. The refined profile parameters are
used for the calculation of integrated peak intensities. IC-POWLS
uses these intensities as observations for magnetic structure
refinement calculations. With our analysis, we confirm that
\chem{Ca_3CoRhO_6} crystallizes in the rhombohedral space group
R\=3c with the lattice constants, \textit{a} = 9.214(2)~\AA \ and
\textit{c} = 10.742 (2)~\AA \ at 293 K, in good agreement with
those from x-ray diffraction~\cite{15}. The atomic site
occupations are \textit{Ca} in 18e~(x,0,1/4), \textit{Co} in
6a~(0,0,1/4), \textit{Rh} in 6b~(0, 0, 0) and \textit{O} in
36f~(x, y, z). As the temperature is lowered below 100~K, a
prominent new line appears in the pattern around 15.5~deg
scattering angle attributable~\cite{16} to the onset of long-range
magnetic ordering. The parameters characterizing the long-range
magnetic order obtained from our analysis
of the diffraction data at 4.2~K
are tabulated in table~\ref{t.1}.
We obtain a moment value of 3.7~(4)~$\mu_B$ parallel to $c$ per
antiferromagnetically ordered Co-ion (4 out of 6 in the unit cell)
in fair agreement with the value of 4~$\mu_B$ for Co-ions in a
trivalent state and with high-field magnetization data at 4.2~K~
and no moment on \chem{Rh}, in accordance with \cite{16}.

\begin{figure}
\onefigure{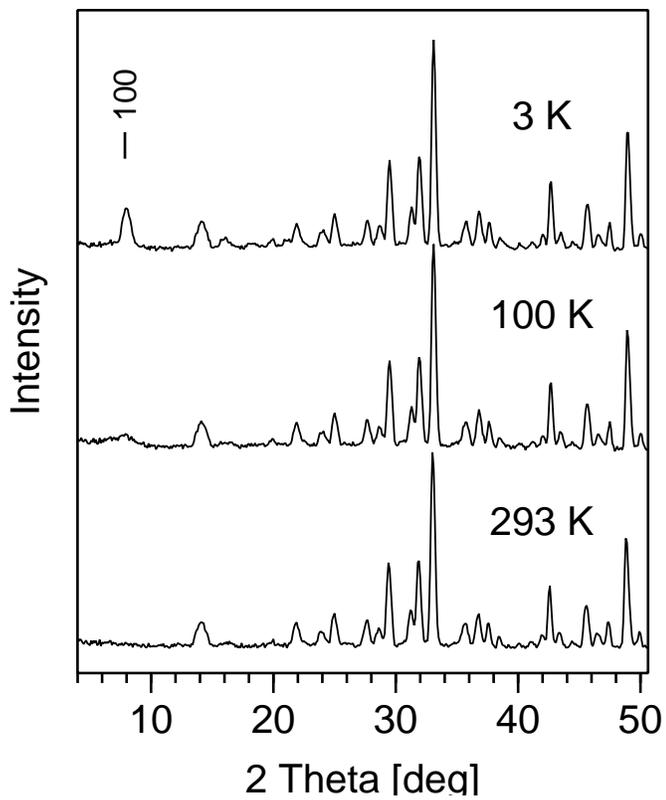} \caption{Comparison of \chem{Ca_3CoRhO_6}
diffraction patterns measured at 3~K, 100~K, and 293~K using a
neutron wavelength of 1.095~\AA. The strongest magnetic reflection
(100) is clearly visible as superstructure reflection at 3~K and
as a diffuse intensity enhancement at 100~K. } \label{f.2a}
\end{figure}

\begin{figure}
\onefigure{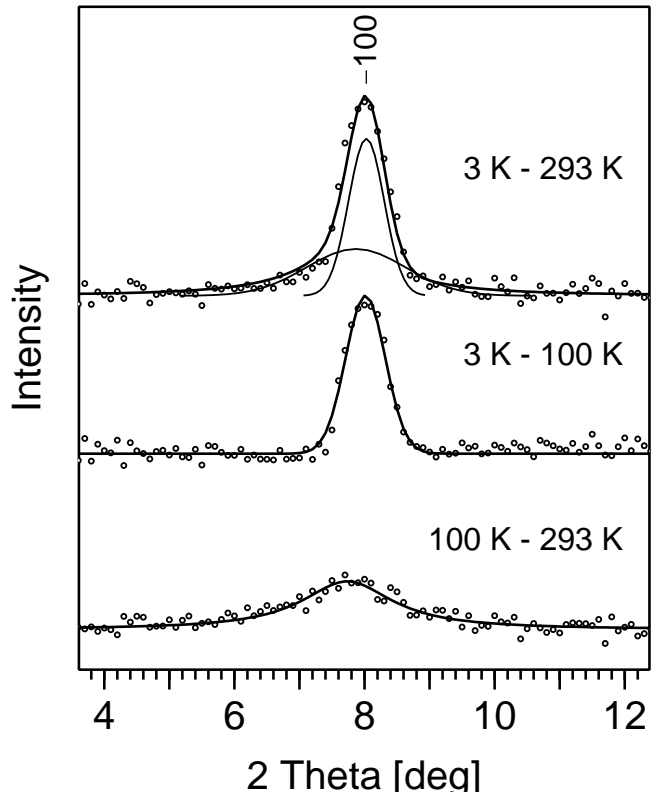} \caption{Magnetic reflection (100) of
fig.~\ref{f.2a} in an expanded  scale according to its appearance
in the temperature difference patterns. Profile fits (lines) show
that the peak at 3 K is composed of a Gaussian (resolution
limited) Bragg peak resulting from long-range magnetic order and a
diffuse share which also exists at 100 K.} \label{f.2b}
\end{figure}

\begin{figure}
\onefigure{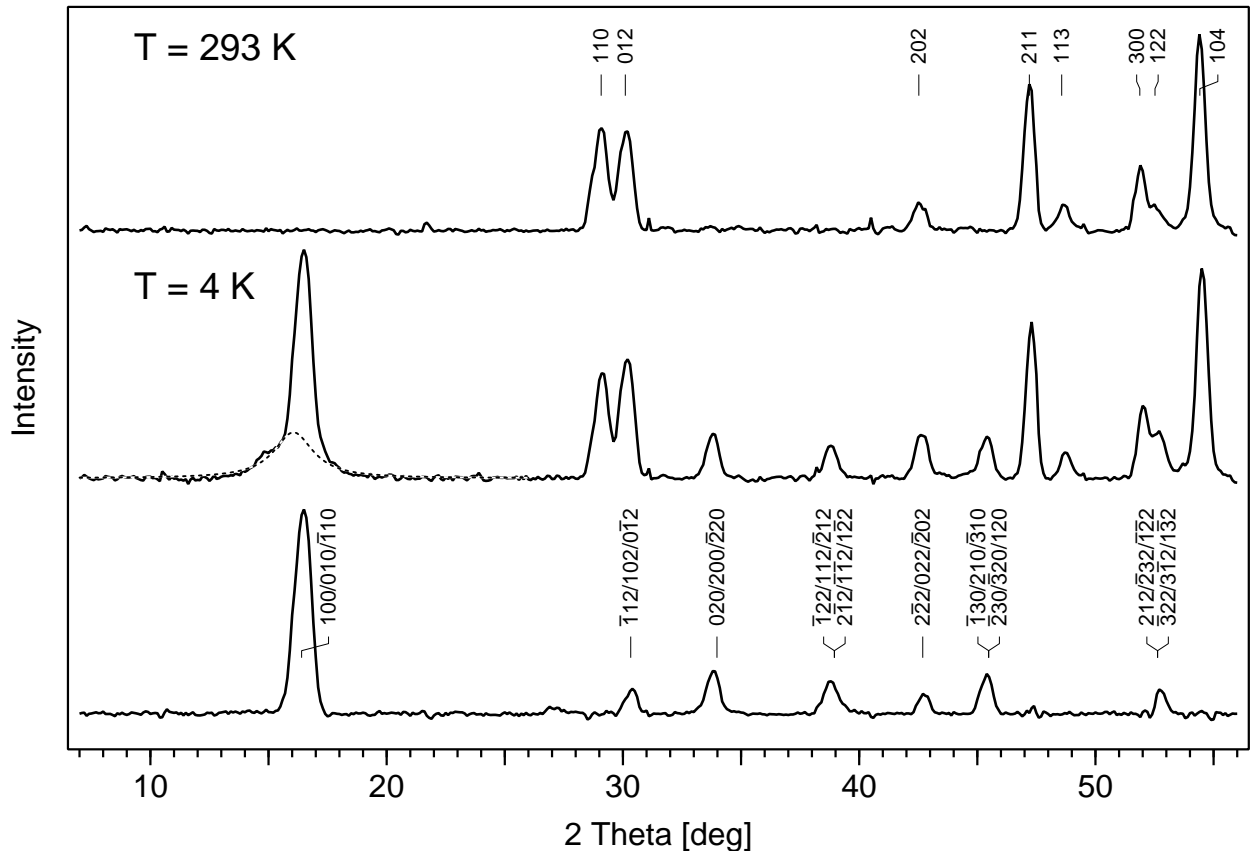} \caption{Comparison of the diffraction
patterns of \chem{Ca_3CoRhO_6} measured with increased resolution
at 293~K (crystal structure indexing) and at 4~K using a neutron
wavelength of 2.332~\AA. Note the clear separation of the (110)
and (012) Bragg reflections if compared to the low resolution data
of fig.~ \ref{f.2a}. The strongest magnetic peak at $2\Theta$ =
16.5~deg is superimposed by a broad diffuse share (dotted line) as
obtained by peak profile fits using two Gaussian-type curves. The
temperature difference pattern (4~K - 293~K, bottom) shows the
pure long-range magnetic Bragg scattering after subtraction of the
diffuse part.} \label{f.2c}
\end{figure}

We now turn to the point of central importance. A careful look at
our raw diffraction patterns shown in fig.~\ref{f.1} indicated to
us that there is in addition to the magnetic Bragg peak a broad
weak peak ("magnetic short-range-order-like") that is present at
all investigated temperatures, both above and below the long-range
magnetic ordering temperature. We have carefully obtained the
patterns at several temperatures for the shorter wavelength
(1.095~\AA) and the typical patterns obtained are shown in fig.~
\ref{f.2a} for 3, 100, and 293~K. The strongest magnetic
reflection (100)  is clearly visible at $2\Theta$ = 8~deg as a
superstructure reflection at 3~K and a diffuse intensity
enhancement at 100~K (figs.~\ref{f.2a} and \ref{f.2b}). This is
visible even in the diffraction patterns measured with increased
resolution using a neutron wavelength of 2.332~\AA (fig.~
\ref{f.2c}). It may be noted that in the latter case there is a
clear separation of the (110) and (012) Bragg reflections when
compared to the low resolution data of fig.~\ref{f.2a} and the
strongest magnetic peak at $2\Theta$ = 16.5~deg is superimposed by
a broad diffuse share (dotted line) as obtained by peak profile
fits using two Gaussian-type curves. In order to highlight the
features due to the diffuse part, we have fitted the curves (for
short wavelength) around $2\Theta$ = 8~deg to a superposition of
two Gaussians with different widths (one of them resolution
limited), as shown in fig.~\ref{f.2b}.

\begin{figure}
\onefigure{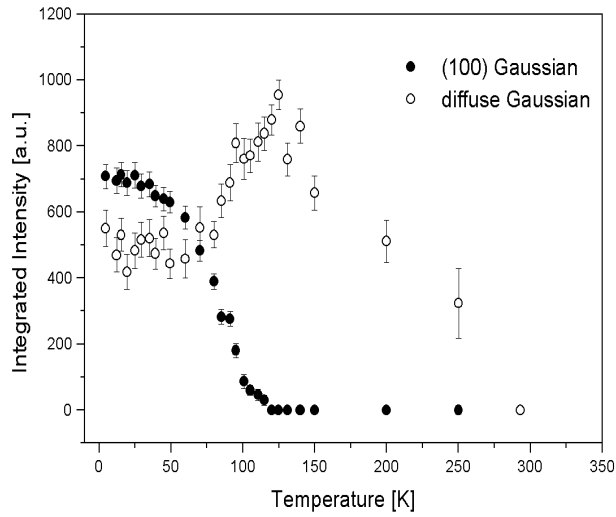} \caption{Temperature dependence of the
integrated (100) magnetic Bragg intensity and of the diffuse
intensity of \chem{Ca_3CoRhO_6} (measured with $\lambda$ =
1.095~\AA) after subtraction of a constant background assuming
Gaussian line shapes for both lines.} \label{f.3a}
\end{figure}

\begin{figure}
\onefigure{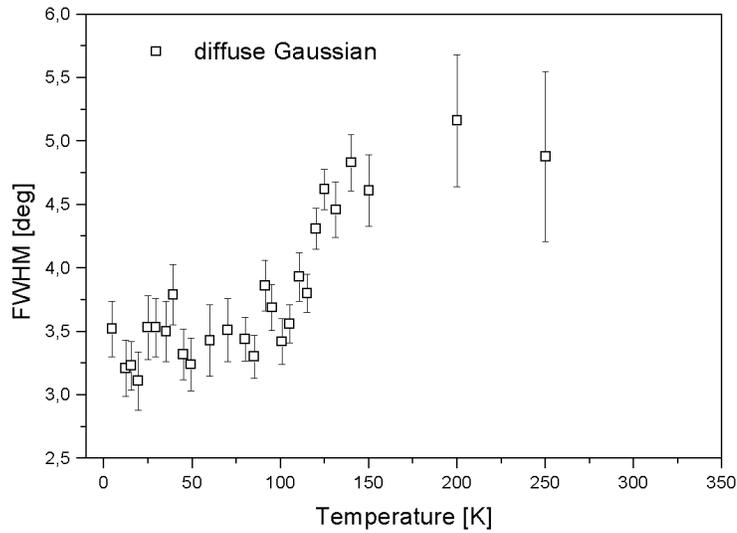} \caption{Temperature dependence of the line
width (\textit{FWHM}) of the diffuse peak of \chem{Ca_3CoRhO_6}.}
\label{f.3b}
\end{figure}

\begin{figure}
\onefigure{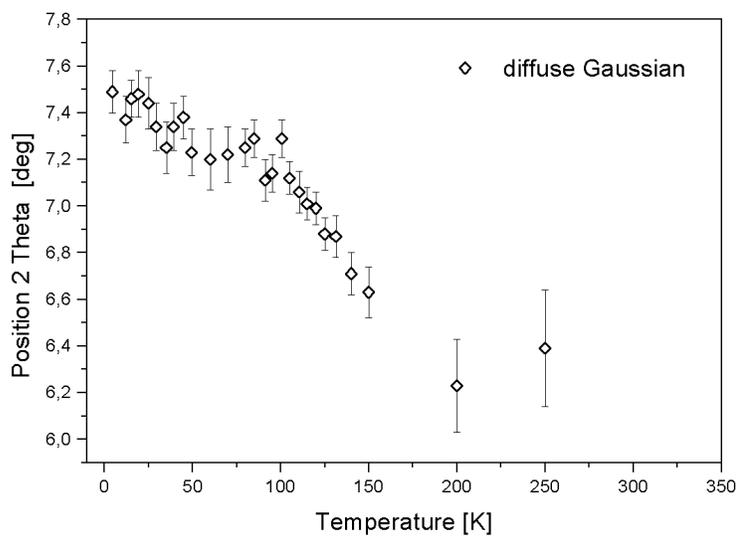} \caption{Temperature dependence of the angular
position of the diffuse peak of \chem{Ca_3CoRhO_6} assuming a
Gaussian line shape.} \label{f.3c}
\end{figure}

In figs.~\ref{f.3a}, \ref{f.3b}, and \ref{f.3c}  we present the
temperature dependence of the relevant parameters (intensity,
width and position, respectively) of the diffuse peak, apart from
showing the temperature dependence of the intensity related to the
long-range magnetic order.

The most remarkable feature is that the diffuse peak intensity
persists in the entire temperature range of investigation as
though the magnetic ions
(Co-3 and Co-6, see table~\ref{t.1})
responsible for this peak are
only weakly coupled to those undergoing long-range magnetic
ordering
(Co-1, Co-2, Co-4, and Co-5).
 The only influence of long-range
magnetic ordering appears to narrow this peak marginally
and to pull out some intensity
below 100~K. A closer inspection of
the temperature dependence of the intensity of the diffuse peak
(fig.~\ref{f.3a}) is quite revealing. The peak intensity increases
as the temperature is lowered below 300~K, attaining a maximum
around 100 to 150~K, followed by a decrease at lower temperatures.
This qualitatively mimics the behavior of $\chi$ expected for
antiferromagnetic correlations in one-dimensional chains~\cite{21}
and is actually observed in the $\chi$ data. The reader may see in
fig.1 of ref.~\cite{15} the tendency for the flattening of $\chi$
in the range 100 - 150~K as though there is a Bonner-Fischer-type
peak and the fall of $\chi$ below 100~K is presumably intercepted
by the enhancement of $\chi$ due to the onset of long range
magnetic order. Thus, the present neutron data is able to
distinctly delineate the features due to long-range magnetic
ordering and the intra-chain effects. We therefore believe that
the so-called "incoherent chains" described in the introduction
retain their individuality ("isolated intra-chain interactions")
down to low temperatures. The width of the broad diffuse peak
underneath the strongest magnetic Bragg reflection (100) (compare
fig.~\ref{f.3b}) has been used for an estimation of the
correlation length at 4~K. The estimation is based on the
reciprocal relation between cluster size \textit{D} [\AA] and peak
half width broadening \textit{FWHM} [deg] according to
\begin{equation}
 D  =  \frac{\lambda\cdot 57.3}{cos \Theta  \cdot FWHM}
\end{equation}
By taking the actual experimental parameters (neutron wavelength
$\lambda$ = 1.096~\AA, central diffuse peak position $2\Theta$ =
7.4~deg and \textit{FWHM} = 2.75~deg after correction for
instrumental resolution), the calculation results in a
characteristic value of 23~\AA \ for the linear extension of the
antiferromagnetic chain segments that coexist with the long-range
magnetic order. In the region without long-range order,
\textit{i.e.} above 100~K, the diffuse peak is somewhat broader,
hence the characteristic length for the one-dimensional cluster
size is correspondingly smaller (16~\AA). Thus, the influence of
long-range magnetic order is to increase the correlation length.
The position-shift of the diffuse peak with decreasing temperature
to larger scattering angles (see fig.~\ref{f.3c}) indicates the
reduction of short-range correlations from larger distances and/or
larger numbers of Co atoms (nearest, next nearest and even more
neighbors) to a smaller number and finally to only Co-3 and Co-6
atoms with an inter-atomic distance of 6.415~\AA. However, because
of the limitation of the diffuse experimental data in Q-space
(only one maximum of the scattering function is observed) the data
cannot be used for the calculation of a pair-correlation function.

\section{Conclusion}
    From a detailed neutron diffraction investigation we find direct
    evidence for the coexistence of  isolated
    intra-chain interactions and long-range magnetic ordering due
    to inter-chain interactions from the rest of the chains in a
    spin-chain system, \chem{Ca_3CoRhO_6}. This is  a fascinating finding
    in the field of quasi one-dimensional magnetism.

\acknowledgments Technical assistance by R. Skowronek in
performing the neutron diffraction experiments is gratefully
acknowledged.

\newpage
\begin{table}
\caption{Parameters of the antiferromagnetic structure of
\chem{Ca_3CoRhO_6} at 4.2~K.} \label{t.1}
\begin{center}
\begin{tabular}{cccc}
magnetic atom&position in chemical cell&moment orientation&magnetic moment [$\mu_B$]\\
Co-1 & 2/3, 1/3, 0.0833 &    +  (parallel to c)     &3.7(4) \\
Co-2 &  0 ,  0 , 0.25   &    -  (anti-parallel to c)& 3.7(4)\\
Co-3 & 1/3, 2/3, 0.4167 &       not ordered         & 0\\
Co-4 & 2/3, 1/3, 0.5833 &    +  (parallel to c)     & 3.7(4)\\
Co-5 &  0 ,  0 , 0.75   &    -  (anti-parallel to c)& 3.7(4)\\
Co-6 &  1/3, 2/3, 0.9167&       not ordered         &0
\end{tabular}
\end{center}
\end{table}
\newpage

\end{document}